\documentclass[twocolumn, trackchanges]{aastex631}

\usepackage{amsmath}
\usepackage{graphicx}
\usepackage{amssymb}
\usepackage{bm}
\usepackage{verbatim}
\usepackage{xcolor}
\usepackage{mathtools}
\usepackage{amsfonts}
\usepackage{graphics}
\usepackage{listings}
\usepackage[normalem]{ulem}      
\usepackage{color,soul}
\newcommand{\EQ}{\begin{equation}}
\newcommand{\EN}{\end{equation}}
\newcommand{\EQA}{\begin{eqnarray}}
\newcommand{\ENA}{\end{eqnarray}}
\newcommand{\eq}[1]{(\ref{#1})}

\newcommand{\Eq}[1]{Eq.~(\ref{#1})}

\newcommand{\Fig}[1]{Fig.~\ref{#1}}

\newcommand{\jdotE}{\int \int \tbf{j}\cdot \tbf{E}\ dx dz}

\newcommand{\gmx}{\gamma_{max}}
\newcommand{\gmn}{\gamma_{min}}
\newcommand{\gmxp}{\gamma_{max, N_0}}
\newcommand{\gmnp}{\gamma_{min, N_0}}
\newcommand{\lb}{\left(}
\newcommand{\rb}{\right)}

\def \ccr {\textcolor{red}}

\def \palvi{\textcolor{magenta}}
\def \tbf{\textbf}

\DeclareMathOperator{\sech}{sech}

\begin{document}

\title{Magnetic Reconnection: An Alternative Explanation of Radio Emission in Galaxy Clusters}

\author[0000-0002-9841-3756]{Subham Ghosh}
\affiliation{International Centre for Theoretical Sciences, 
Bangalore, 560089, India}

\email{subham.ghosh@icts.res.in}


\author{Pallavi Bhat}
\affiliation{International Centre for Theoretical Sciences, Bangalore, 560089, India}
\email{pallavi.bhat@icts.res.in}


\begin{abstract}
Observations of galaxy clusters show radio emission extended over almost the system scale, necessitating mechanisms for particle acceleration. 
Previous models for acceleration such as diffusive shock acceleration and that due to turbulence {can} fall short in terms of efficiency. 
In this letter, we propose the possibility of acceleration via magnetic reconnection. In particular, we invoke the plasmoid instability which has been previously applied to understand particle energization in high energy systems. 
Turbulence in galaxy clusters lead to fluctuation dynamos that are known to generate magnetic fields structures consisting of sharp reversals. 
These form natural sites of reconnection. We perform Particle-In-Cell (PIC) simulations of the plasmoid instability in collisionless and nonrelativistic plasmas. 
We show that the resulting {electron} energy spectra have power law indices that are consistent with that inferred from observations. 
{Our estimates show that the acceleration timescales are much smaller than the lifetime of the reconnecting magnetic structures indicating the feasibility of our model. 
The synchrotron radio luminosity estimate is about   $10^{41}$ ergs/s, agreeing with observations.} 
{Finally, we find that the maximum achievable Lorentz factor can go upto} $10^5$ indicating that acceleration due magnetic reconnection is a promising avenue for understanding the origin of nonthermal emission in galaxy clusters.
\end{abstract}

\keywords{Galaxy clusters (584) --- magnetic reconnection (1504) ---Cosmic rays (329) --- Intracluster medium (858) --- High energy astrophysics (739)}

\section{Introduction} \label{sec:intro}

In galaxy clusters, the thermal plasma, with temperatures ranging from $10^7$ to $10^8$ Kelvin, emits soft X-rays. 
{Additionally, in many systems, diffuse extended radio emission is observed \citep{Giovannini_2009, Bonafede_2015, vanWeeren_2019, Cuciti_2021}. These radio emitting regions are categorized in two groups: radio halo and radio relics.  Radio halo is observed in the central volume of the clusters, typically, within a distance of 1Mpc. However, radio relics are found primarily in the periphery of the clusters. We focus more on the radio halos in this letter.}  
This radio emission is a signature of the non-thermal component of the intracluster medium {(ICM)}. 
Haloes invariably exhibit steep spectra with the spectral index $\alpha \gtrsim 1$ \citep{Feretti_2012}, where $\alpha$ is defined as $S_{\nu}\propto \nu^{-\alpha}$ for the luminosity flux density $S_{\nu}$ at the frequency $\nu$. The nonthermal emission is necessarily linked to the presence of relavistic particles in the plasma. 
 
The primary argument necessitating electron re-acceleration is that the radiative lifetime of nonthermal electrons due to energy loss from synchrotron and inverse Compton processes, is approximately $10^8$ years \citep{Sarazin_1999}. The diffusion velocity of the electron population, which is of the order of the Alfv\'en speed (approximately $100$ km/s; \citealt{Feretti_2012}), and the corresponding diffusion timescale of approximately $10^{10}$ years, it becomes evident that the radiative timescale alone is insufficient for particles to diffuse through the cluster volume and emit radio frequencies. Therefore, an additional re-acceleration process is required. 

The mechanisms for re-acceleration that have been studied previously include diffusive shock acceleration (DSA; \citealt{Blandford_1978, Drury_1983, Ensslin_1998, Hoeft_2007}) and acceleration due to turbulence \citep{Brunetti_2001, Petrosian_2001, Brunetti_2007, Brunetti_2016}. 
DSA is expected to elucidate particle acceleration in radio relics i.e. in the outer regions of galaxy clusters where turbulence {can be} supersonic \citep{Brunetti_2014, Paola_2021}, though it is at odds with observational data \citep{Vazza_2014, Vazza_2015}. Morover, it falls short in the central cluster regions, where turbulence is predominantly subsonic \citep{Ryu_2008, Porter_2015, Roh_2019}.
On the other hand, acceleration due to turbulence is a Fermi second-order process and is known to be not very efficient \citep{Ensslin_2011}. 
In particular, these processes have been also studied in collisionless plasmas as suitable for galaxy clusters \citep{Caprioli_2014, Guo_2014}.
In this letter, we introduce a mechanism previously unexplored in the context of galaxy clusters known as magnetic reconnection. In this process, magnetic fields undergo topological reconfiguration, due to reconnection of anti-parallel field lines at sites with large current density, accompanied by energy conversion from magnetic to kinetic, heat and particle acceleration.

{Magnetic reconnection has been explored extensively in high energy contexts of pulsar wind nebulae \citep{Cerutti_2020, Lu_2021} and AGN jets \citep{Sironi_2014_2, Zhang_2023}, for collisionless plasmas, in a relativistic avatar \citep{Guo_et_al_2015, Werner_2018}}. 
These studies have shown the possibility of obtaining hard spectra {of particle energy distribution $f(\varepsilon)\propto \varepsilon^{-p}$} with a particle energy spectral index as small as $p=2$ where $p=2\alpha + 1$. 
In this letter, we show that with magnetic reconnection mechanism in nonrelativistic plasmas such as that in galaxy clusters, it is possible to accelerate particles efficiently to obtain nonthermal emission with a spectral index range of $\alpha \sim 1 - 3$.
Previous studies that explored reconnection in galaxy clusters \citep{Brunetti_2016}, did so with models that differ significantly from our approach. {These previous approaches were applied to other high energy contexts as well \citep{dalPino_2005, delValle_2016}}. 

For magnetic reconnection to take place, we need sites in plasma that have large current densities as a consequence of the interaction of reversing magnetic fields. Such sites of reconnection naturally occur in magnetic fields generated by fluctuation dynamo  \citep{Galishnikova_2022, Beattie_2024}. In the absence of mean rotation, only fluctuation dynamo or small-scale dynamo can operate in galaxy cluster turbulence \citep{Subramanian_2006, Bhat_2013}, which is known to produce fields that are intermittent. In the kinematic regime, the fields are largely dominated by magnetic structures which are reversing on resistive scales \citep{Schekochihin_2002}. However such structures diminish by saturation and fields become more coherent \citep{Bhat_2013}. Nonetheless, even in saturation, it is expected that the dynamo generates intermittent regions of reversing fields with extreme strengths \citep{Brandenburg_2005}. 

Reversing field structures are unstable to tearing mode instability \citep{Zweibel_2009}. 
At large Lundquist numbers (or equivalently large magnetic Reynolds numbers), tearing modes lead to what is known as the plasmoid instability, which has been discovered to lead to fast reconnection rates \citep{Loureiro_2007, Bhattacharjee_2009}. In this letter, we demonstrate using particle-in-cell (PIC) simulations that plasmoid instability in collisionless nonrelativistic plasmas leads to efficient particle acceleration. 

\begin{figure*}
\includegraphics[width=\linewidth]{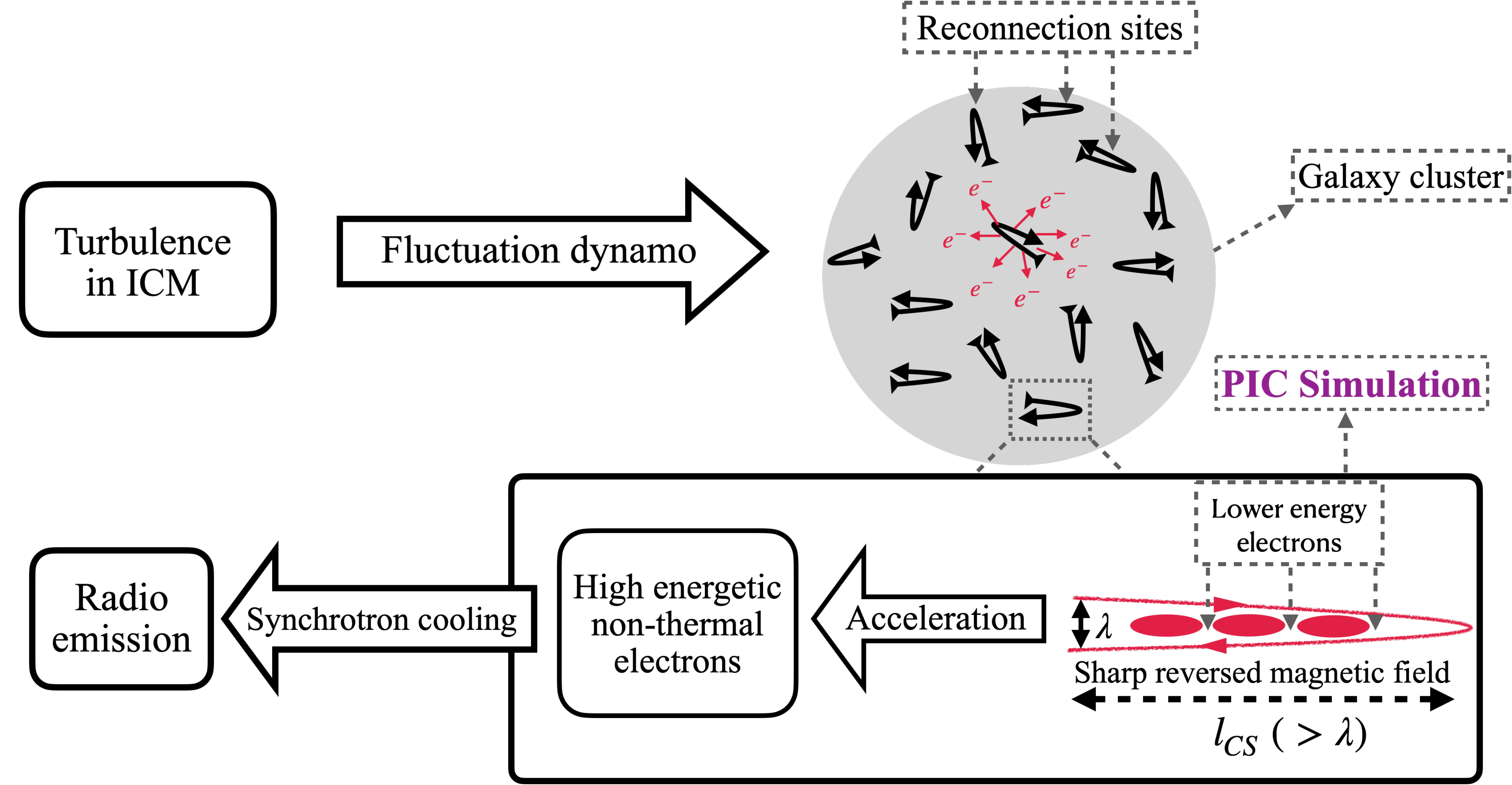}
\caption{\label{fig:bigger_picture} Flow chart of our model. This shows how the turbulent energy in the ICM gets converted into radio. The fluctuation dynamo converts a part of the kinetic energy of turbulence to magnetic energy, which in turn gets converted to particles energy via magnetic reconnection. Eventually, the particles energy gets converted to radio.} 
\end{figure*}

With the observed synchrotron emission frequencies of $\sim 0.1$GHz \citep{Pandey-Pommier_2013} to $\sim 1.4$GHz \citep{Feretti_2012, Paul_2023} and the typical magnetic field strength of a few to 10s of $\mu G$, the Lorentz factor associated with nonthermal electrons can be computed to be in the range of $\gamma\sim 10^3 - 10^5$. 
We find that it is likely possible to explain such large Lorentz factors as well with magnetic reconnection as the re-acceleration process. 
{Apart from this, the synchrotron radio luminosity observed in the radio halo, can also be explained from the energization of the electrons in the magnetic reconnection process.} 

{A flowchart of our model has been shown in a schematic diagram in \Fig{fig:bigger_picture}. Schematically, it shows that the turbulence in the ICM leads to a fluctuation dynamo which generates, as previously mentioned, intermittent reversing strong field structures naturally.
These structures are unstable to reconnection and in the process, accelerate particles.
Thus, these regions could act as reservoirs of high energetic electrons. We consider one such reconnection site and perform numerical simulations of reconnection using a PIC code as indicated in the \Fig{fig:bigger_picture}. 
We observe from the simulation that after participating in the magnetic reconnection process, low energy electrons evolve to a population of relativistic and non-thermal electrons. When these electrons cool through synchrotron, they emit in radio. }

This paper has been organized as follows. In section \ref{sec:setup}, we describe the details of the PIC simulations. The simulation results are described in section \ref{sec:Numerical Results}. The application of the simulation results are described in section \ref{sec:Application to galaxy clusters}. We, then, conclude in section \ref{sec:Discussion and Conclusion}.

\section{simulation setup} 
\label{sec:setup}
We use a publicly available PIC code \textit{WarpX} \citep{warpx_cite}. Our simulation domain is two-dimensional ($x \times z$). The plasma consists of electrons and positrons. The plasma component corresponding to the current sheet where the reconnection takes place is localised to the central part of the domain and is initialized to a temperature of $T_e=T_i=T$ (this corresponds to galaxy cluster temperatures).  
Another component of the plasma, whose acceleration we would like to study, is spread out through the domain and is at a lower temperature $T_b \ll T$; we call this the background plasma. 
The plasma component at temperature $T$ also is initialized with bulk velocities along $y$-direction in opposite directions. The arising drift velocity between electrons and positrons leads to a current and is consistent with an equilibrium magnetic field corresponding to the Harris sheet \citep{Harris_1962}, 
$B_x = B_0 \tanh\left({z}/{\lambda}\right)$, 
where $\lambda$ is the half-width of the initial current sheet. The corresponding 
number density is given by
$n = n_0\ {\sech}^2\left({z}/{\lambda}\right)$, {where $n_0$ is the peak density of the drifting particles}.
The magnetic pressure of equilibrium field is balanced by thermal pressure of the plasma, 
\begin{equation}
    \frac{B_0^2}{2\mu_0} = 2n_0K_BT= 2n_0\theta mc^2,
    \label{eq:energy_balance}
\end{equation}
where, $\theta = K_BT/mc^2$ is the dimensionless temperature. When $\theta < 0.1$, we are in nonrelatistic regime and the plasma is intialized using a Maxwell-Boltzmann distribution. For $\theta \ge 0.1$, we use Maxwell-Juttner distribution. While we have run simulations for different values of $\theta$, the fiducial value for $\theta$ is $0.05$ corresponding to a temperature of $\sim 10^8$K, as for galaxy clusters. 
{Note that \Eq{eq:energy_balance} implies a plasma beta of approximately 1, whereas in galaxy clusters, the plasma beta is typically closer to 100. 
However, as mentioned in section~\ref{sec:intro}, our focus is on regions with strong reversing magnetic fields generated by the fluctuation dynamo, where the plasma beta can be significantly lower. The current setup is primarily intended as an experimental framework. That said, it would be possible to explore a wider range of plasma beta values by adopting a force-free configuration instead of one based on pressure balance.}

The background plasma at lower temperature $T_b$ has a constant density $n_b = 0.1 n_0$. We add a perturbation to the equilibrium magnetic field of the form in \citet{Daughton_2007, Ricci_2004}, 
such that it 
produces an initial $x$-point at the center of the simulation domain. 
The fastest growing mode of the tearing instability theory is the largest possible in the domain and thus a single island arises along the length of the current sheet as expected. 

The domain sizes are given by, $-L_x\leq x \leq L_x$ and $-L_z\leq z \leq L_z$, respectively, with $L_x \times L_z = 204.8d_e \times 102.4 d_e$ where $d_e$ is the plasma skin depth. It is defined as $d_e= c/\omega_p$, with $\omega_p$ as the plasma frequency and given by $\omega_p  = \sqrt{{n_0e^2}/{m\epsilon_0}}$, {where $m$ is the mass of an electron and $\epsilon_0$ is the free space permittivity}. The initial half-width of the current sheet is $\lambda = 4d_e$. The resolution is $N_x\times N_z = 2048\times 1024$. 
We define the magnetization w.r.t the plasma associated with current sheet as $\sigma_{cs} = B_0^2/\mu_0mc^2n_0 = (\Omega_c/\omega_p)^2 = 4\theta$ as opposed to the background particles. When $\theta < 0.1$ and $\sigma_{cs} < 0.4$, the reconnection is nonrelativistic. 

The simulations have 128 particles per cell for both components of the plasma. The boundary conditions along $x$-direction for both field and particles are periodic. However, along $z$-direction, the boundary conditions are reflecting and perfectly conducting for the particles and fields, respectively.  

\section{Numerical Results}
\label{sec:Numerical Results}
\begin{figure*}
\centering
\includegraphics[width=\textwidth]{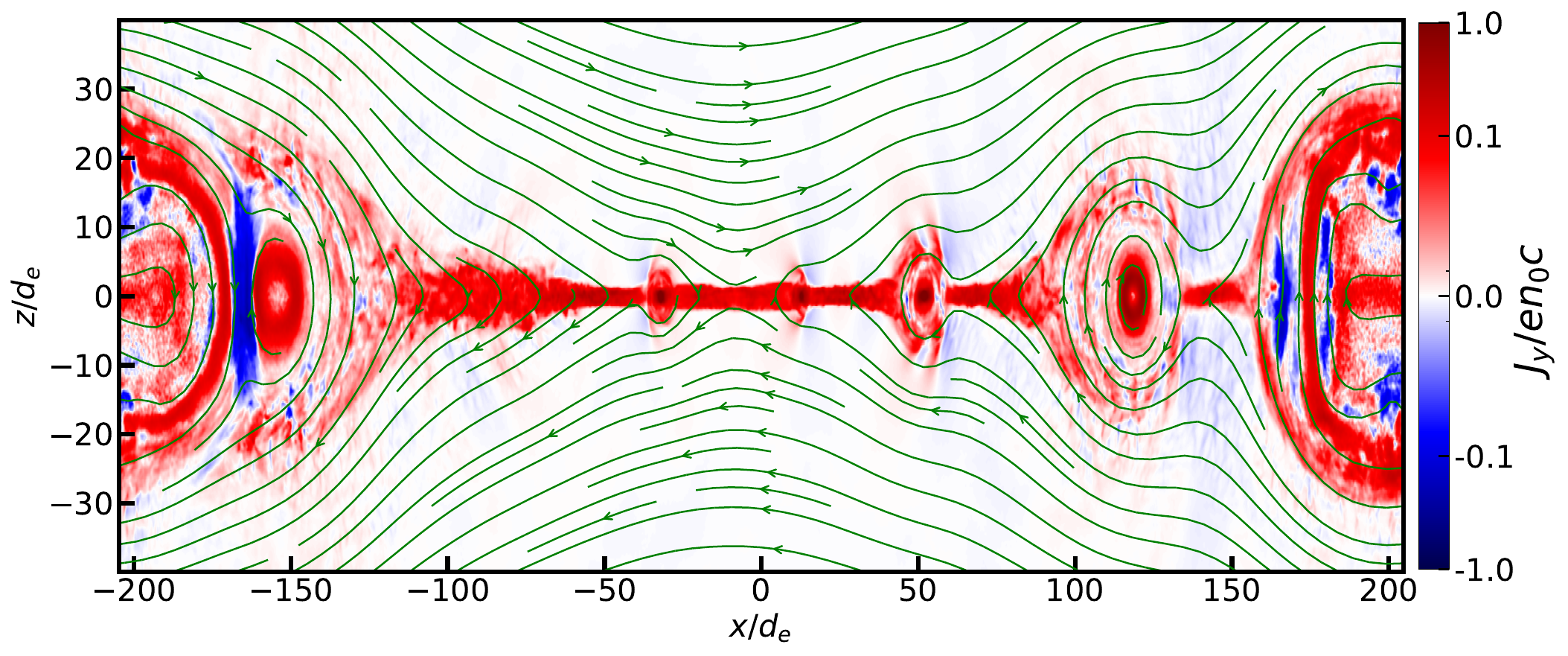}
\caption{\label{fig:jy_iteration_theta_point05_n_1000_single} Surface plots of $j_y$ at $\omega_p t = 769.33$ from the run with $\theta=0.05$. The in-plane ($x-z$) magnetic field (green lines) is overlaid on top of $J_y$ with the green lines. {The entire $z$-axis has not been plotted here. Only the significant range of it is shown here for better presentation.}}
\end{figure*}

\begin{figure}
\includegraphics[width=\linewidth]{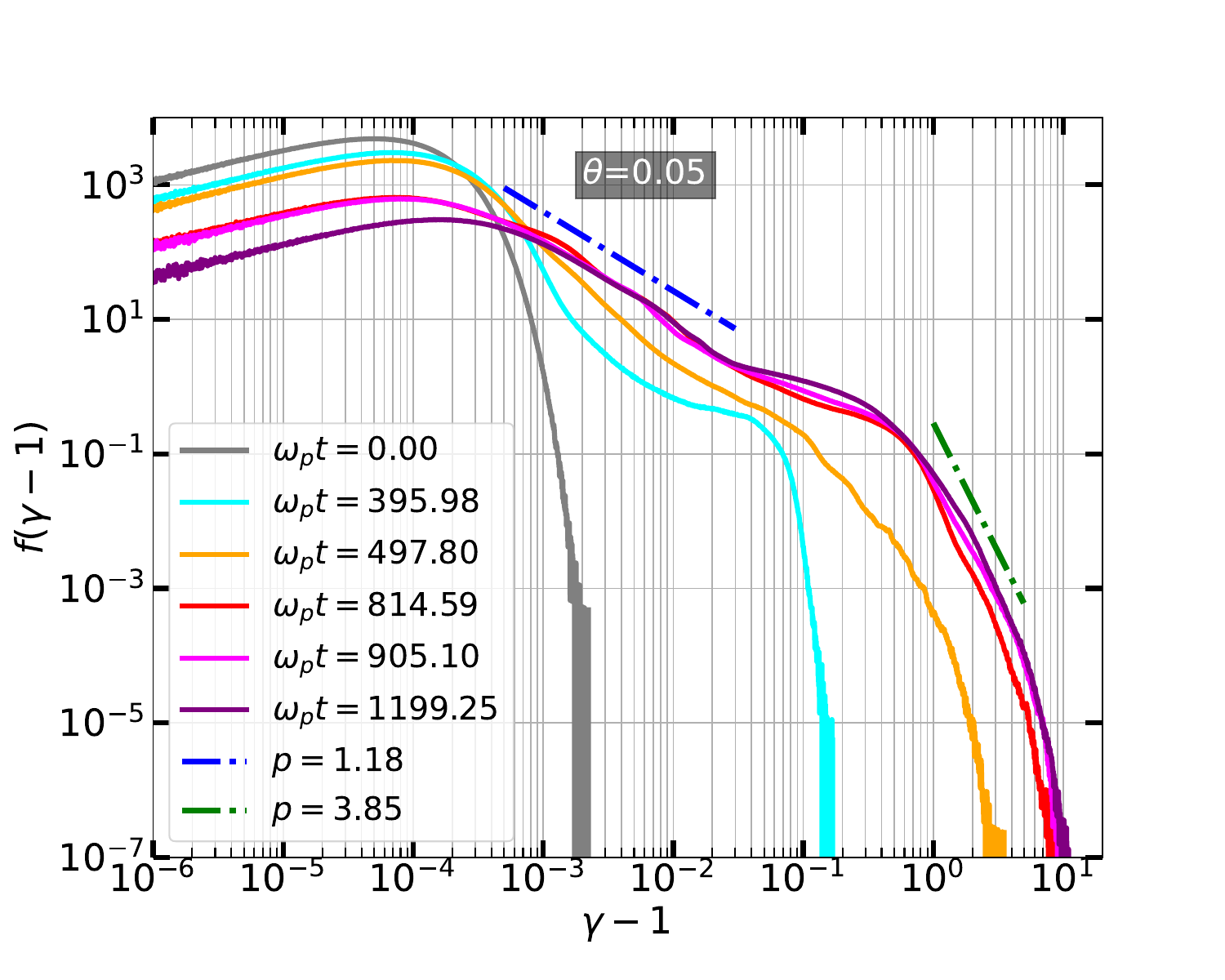}
\caption{\label{fig:energy_spectra_theta_point05} The distribution of background electrons' kinetic energy per unit rest mass energy for run with $\theta = 0.05$ and its evolution with time. The blue and green dashed-dotted lines indicate the slopes in the spectra {corresponding to largely suprathermal and ultra-relativistic ranges, respectively}.}
\end{figure}

\begin{figure}
\includegraphics[width=\linewidth]{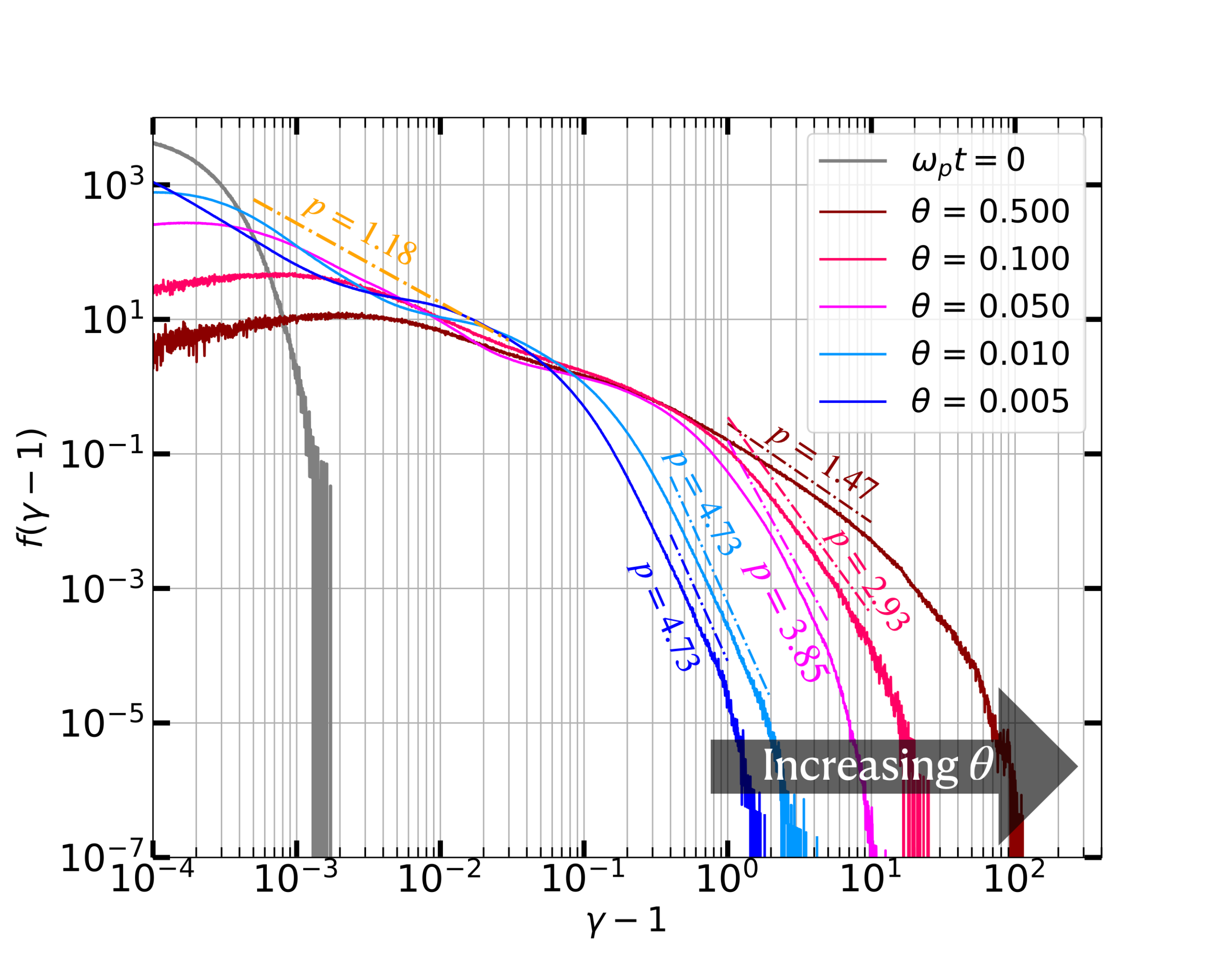}
\caption{\label{fig:hist_ke_ke_rel_nonrel_diff_theta_ylim} The distribution of background electrons' kinetic energy per unit rest mass energy for different $\theta$ at the final stage of its evolution.} 
\end{figure}

Magnetic reconnection is a phenomenon that breaks frozen-in flux condition due to non-ideal effects. While in MHD, the flux freezing breaks due to microscopic resitivity (that arises due to the collisions between electrons and ion as they drift past each other), in collisionless plasmas as we have considered here, the magnetic fields slip due to electron inertia. 
The Harris sheet equilibrium, on perturbation, leads to tearing mode instability, which involves reconnection of magnetic fields at the X-point giving rise to a single island first \citep{FKR_1963, Coppi_1976, Drake_1977}. 
Eventually, the $X$-point collapses to generate secondary current sheets, which produce small secondary islands or plasmoids \citep{Syrovatskii_1971, Loureiro_2005}. 
In \Fig{fig:jy_iteration_theta_point05_n_1000_single} we show that plasmoids,  begin to form around $x/d_e = 0$ and due to the ram pressure of the flow, move towards the boundaries of the domain while growing in size.   

The growing plasmoids eventually merge with the primary tearing mode island at the boundary of the simulation domain. 
When the large plasmoids coalesce with the primary island, a current sheet forms at the interface and is perpendicular to the primary current sheet. Such a perpendicular current sheet is observed between $x/d_e =-150$ and $-200$ in \Fig{fig:jy_iteration_theta_point05_n_1000_single}, and also at the other end. The electric field that is generated due to the reconnection between the islands opposes the electric field generated in the primary reconnection. 
Reconnection and plasmoid generation continues  until the magnetic flux is exahusted and all the islands merge leaving a single island in the simulation domain towards the end of the simulation. In this process, the particles interact with the current sheets and islands and get energised.

The background particles have a thermal distribution initially. They get accelerated due to the magnetic reconnections and result in a non-thermal distribution by the end of the simulation. The evolution of energy spectra of the background electrons has been shown in \Fig{fig:energy_spectra_theta_point05}. For a non-thermal distribution of particles with the distribution function $f(\gamma -1)\bm{\propto (\gamma-1)^{-p}}$, {where $\gamma$ is the Lorentz factor}, the power law index, $p$ is defined as 
$p = - {d \ln{f(\gamma -1)}}/{d\ln{(\gamma -1)}}$.
The final saturated particle energy spectrum shows two distinctive slopes. The value of $p$ for first of these two slopes is around $-1$. This has been previously observed in all reconnection related energised particle spectra studies \citep{Guo_et_al_2015, Werner_2018}. {Note that this slope pertains to the suprathermal electrons}.
However the spectrum relevant to non-thermal emission is the second one, which governs largely the ultra-relativistic particles. 

{From \Fig{fig:energy_spectra_theta_point05}, one might mistakenly infer that the shifting of the particle distribution spectrum to the right, with the peak moving to larger values of \( \gamma \), indicates an increase in the temperature of the Maxwellian and suggests thermal X-ray brightening. However, we caution the reader that \Fig{fig:energy_spectra_theta_point05} represents only the background electrons. The primary Maxwellian corresponding to the current sheet (not shown in the paper) remains unchanged in its peak position but develops a tail with slopes matching those observed for the background electron population.}

\begin{figure}
    \centering
\includegraphics[width=\columnwidth]{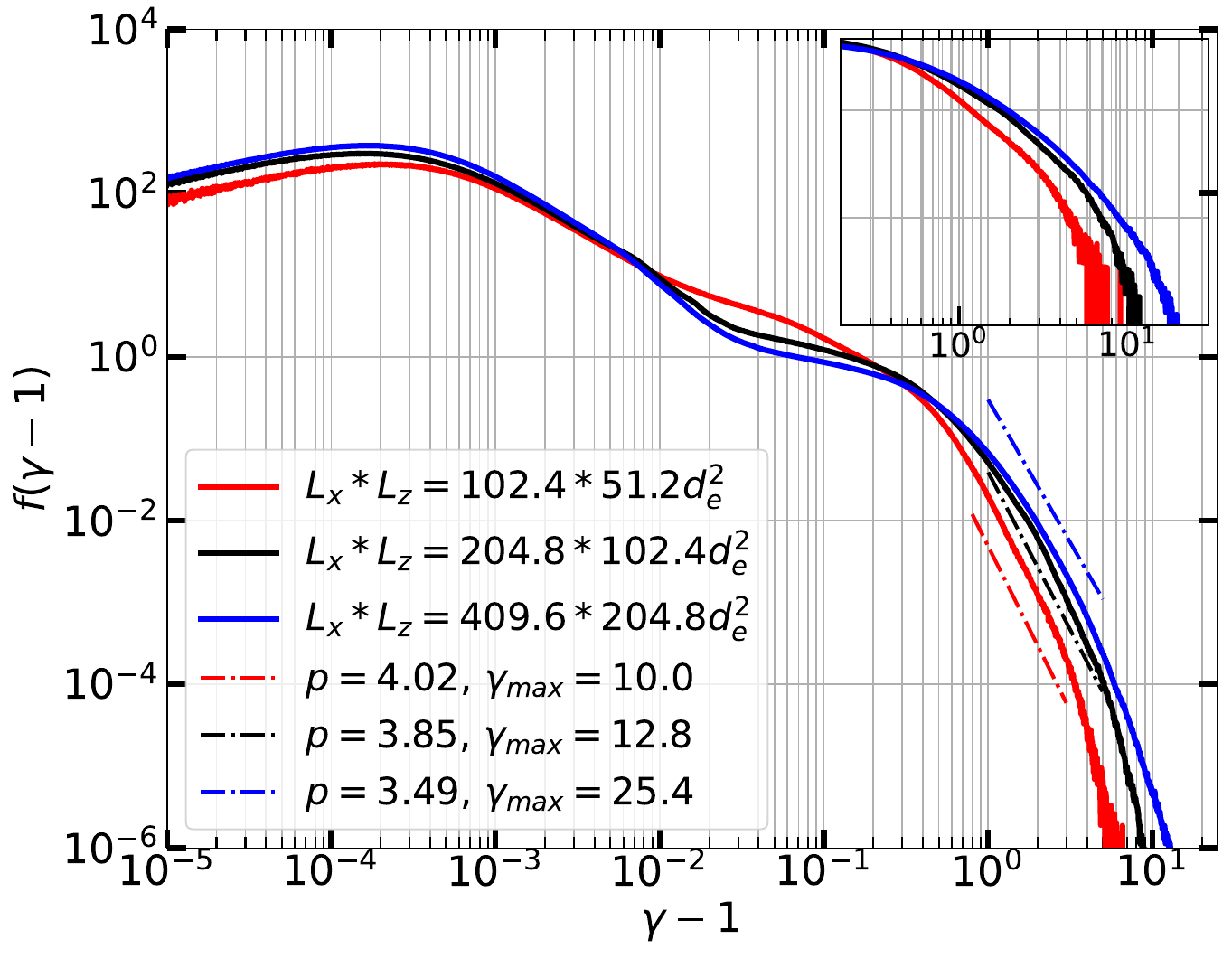}
    \caption{Electron distribution for different box sizes with the same temperature, $\theta = 0.05$. Note that $\gmx$ increases as box
    sizes are getting larger. At the same time, the slopes of the non-thermal distribution are getting shallower comparatively. \label{fig:energy_spectra_theta_point05_diff_boxsizes_inset}}
\end{figure}

In Figure~\ref{fig:hist_ke_ke_rel_nonrel_diff_theta_ylim}, the final particle energy spectra from different simulations with different $\theta$ are shown. The run with fiducial value of $\theta=0.05$ has a slope of $p\sim 3.8$ for the nonthermal particles. 
As $\theta$ increases, the spectrum gets shallower {i.e. $p$ drops to 2.9 from 3.8 when $\theta$ is increased by just a factor of 2, thus the slope is very sensitive to $\theta$.
{Also, this slope value of $p\sim 3.8$ at $\theta=0.05$ should not be considered the final value at this $\theta$, as it varies with the box size. To illustrate this, we have conducted simulations with different domain sizes while maintaining equivalent resolution. We show in \Fig{fig:energy_spectra_theta_point05_diff_boxsizes_inset}, that as the domain size (or equivalently, the current sheet size) is increased, the corresponding slope of the spectrum gets shallower.}  
{We also notice that for larger and larger current sheets, the maximum value of $\gamma$ achievable distinctively increases (can go up to $10^5$ as we show later in section \ref{sec:Maximum acceleration}) and as a result, the corresponding slope value decreases as well. Similar behaviour was already reported in  \cite{Guo_et_al_2015}}. However, it is possible that the value of $p$ can asymptote to a final value with increasing current sheet length.}
In the runs with $\theta > 0.05$, the plasma is relativistic and the acceleration leads to energised particles with higher and higher values of $\gamma$. 

{From Figure \ref{fig:hist_ke_ke_rel_nonrel_diff_theta_ylim}, we notice that as the temperature ($\theta$) is increased, the corresponding maximum energy of the electrons  also increases and the spectra get shallower. To understand this, we refer to equation (\ref{eq:energy_balance}) which shows that $\theta$ directly relates to the available magnetic energy density for magnetic reconnection. Increasing $\theta$ with the other parameters fixed, therefore, implies the increase in available magnetic energy. As a result of that, electrons will have larger kinetic energy for larger $\theta$. Hence, their maximum energy will increase. 
However, the total number of electrons and hence the area under each distribution are same in all the simulations with different temperatures. 
To keep the area under the distribution same for higher $\theta$ with higher energies for electrons, the distribution has to become shallower.}

{We find that the agent for the particle acceleration is the reconnecting electric field, $E_{rec}=E_y$ as we do not have any guide field in the simulation \citep{BB_2007, Zenitani_2001}. See Appendix (\ref{sec:Particle_acceleration_mechanism}) for details. We do not address the particle acceleration mechnanism in greater detail as it is outside the scope of this paper.}

\section{Application to galaxy clusters}
\label{sec:Application to galaxy clusters}

Next, we apply the results from our PIC simulations to assess the re-acceleration 
of electrons in galaxy clusters due to reconnection. If the acceleration timescale is very short compared 
to the radiative cooling time scale, the particle energy spectra we have obtained in our simulations are directly applicable. We show in the next subsection that this is indeed the case. 

The magnetic fields at these reconnection sites need to be sufficiently strong to 
cause efficient re-acceleration. As previously mentioned, 
fluctuation dynamo operates in galaxy clusters due to the plasma turbulence. 
This turbulent dynamo amplifies the field via repeated random stretching  which leads to intermittent regions of high strength fields with sharp reversals. 
Frequency of such regions is higher in kinematic regime when the field 
is still weak, but decreases by saturation. 
Regardless, such strong and reversing magnetic fields can generate large current density structures  that serve as reconnection sites.

We estimate the field strength needed at reconnection sites in the following manner. The plasma beta for galaxy clusters is given by $\beta_g=P_g/(B_g^2/2\mu_0)$ where $B_g$ is the globally averaged magnetic field. In our simulation, we assume initial pressure balance between magnetic and thermal forces (as in \Eq{eq:energy_balance}) before the instability ensues, which sets the field strength required to achieve a certain final particle energy spectrum. If we assume this local thermal pressure strength is similar to the globally averaged value, then we can relate the magnetic field strength at the reconnection site, $B_{rec}$ to that of the global field, $B_g$, 
as $B_{rec} = \sqrt{\beta_g} B_g$.

\subsection{Acceleration efficiency}
\label{sec:Acceleration efficiency}
For efficient re-acceleration, we require that the timescale for acceleration be much smaller than the timescale on which the dynamical magnetic structures (along with the current sheet) survives in this turbulent medium. 
The re-acceleration timescale $\tau_{acc}$ is the same as the magnetic reconnection time-scale, $\tau_{rec}$. 
We make timescale estimates in both fluid MHD regime and in the kinetic regime to get a sense of how the numbers compare. 
It has been well-established empirically (also observed in our simulation) that in the collisionless case, 
$\tau_{rec} \sim 10~\tau_{A_{CS}}$ \citep{Comisso_2016}, where $\tau_{A_{CS}}$ is the Alfv\'en timescale corresponding to the fields around the current sheet. In the MHD case, $\tau_{rec} \sim 100~\tau_{A_{CS}}$ \citep{Bhattacharjee_2009}.
Such an estimate of $\tau_{rec}$ 
sensitively depends on the size of the 
current sheets where the 
reconnections take place, $l_{CS}$. Since the reconnection sites arise at the interface of reversing magnetic fields generated by the fluctuation dynamo, such structures necessarily are  
anisotropic. The longer length scale of such a magnetic structure also governs the length of the current sheet and thus is given by $l_{CS}$. 
Since the system 
is turbulent, these magnetic structures also have a finite lifetime dictated by the eddy turnover time, $\tau_{life} \sim \lambda/v_{CS}$, where $\lambda$ is the width of the magnetic structure with $\lambda < l_{CS}$ and $v_{CS}$ is the flow velocity on that scale. 
The Alfv\'enic timescale $\tau_{A_{CS}}=l_{CS}/v_{A_{CS}}$, where $v_{A_{CS}}$ is the Alfv\'en velocity associated with the reconnection site and thus $v_{A_{CS}}=B_{rec}/\sqrt{\mu_0\rho}$.
We have, 
\EQ
\frac{\tau_{rec}}{\tau_{life}} \sim R^{-1} \frac{l_{CS}}{\lambda}\frac{v_{CS}}{v_{A_{CS}}}
\label{tratio}
\EN
where $R$ is the dimensionless reconnection rate for plasmoid instability with the value of $0.01$ 
in the MHD case and $0.1$ in collisionless case. 
We require the ratio of $\tau_{rec}/\tau_{life} \ll 1$ for reconnection events in this turbulent system to re-accelerate particles efficiently. Next, we have to determine $v_{CS}$ and to do so, 
one could use Kolmogorov-like arguments to relate the velocities on a given scale to the scale size 
by a power-law, $v_{CS} \propto (l_{CS})^s$, where $s=1/3$ for the standard hydrodynamic turbulence. Thus we obtain,  
\EQ
v_{CS}=\frac{B_g}{f_{sat}\sqrt{\mu_0\rho}} \left(\frac{l_{CS}}{L_{out}}\right)^{s}
\label{kolm}
\EN
where, we have taken the saturated magnetic energy of the fluctuation dynamo to be typically a fraction of the kinetic energy \citep{Bhat_2013}, such that $B_g/\sqrt{\mu_0 \rho}=f_{sat} v_{out}$ and $v_{out}$ is the velocity at the outer eddy scale. Going forward, we will suppress the factor of $\sqrt{\mu_0 \rho}$ as it will get cancelled out.  
Considering $S_{CS}=B_{rec} l_{CS}/\eta$ in \Eq{tratio}, we obtain, 
\EQ
\frac{v_{CS}}{v_{A_{CS}}} = \frac{1}{f_{sat}} \left(\frac{B_g}{B_{rec}}\right)^{s+1} \left(\frac{S_{CS}}{S}\right)^{s} 
\label{uratio}
\EN
where $S=B_g L_{out}/\eta$.
Thus, for the timescale ratio, we have, 
\EQ
\frac{\tau_{acc}}{\tau_{life}} = \frac{100}{f_{sat}} \frac{l_{CS}}{\lambda} \left(\frac{B_g}{B_{rec}}\right)^{s+1} \left(\frac{S_{CS}}{S}\right)^{s}.
\label{tratioupd}
\EN
With this result, one may be tempted to conclude that since $S$ is expected to be very large, the timescale ratio of $\tau_{acc}/\tau_{life}$ should be necessarily very small. But firstly only the Spitzer estimate of $S$ is known to be very large and it may not be so in reality. Second, the power law index is a fraction and thus limits the estimate from dropping down to arbitrarily small values.

In the saturated state of the fluctuation dynamo, on any given scale, the 
magnetic tension due to the structures is counterbalanced by the stretching of the fields due to the flow. In absence of a concrete theory for dynamo saturation, we will assume a scale-independent scaling of $\lambda/l_{CS} \sim B_g^2/ v_{out}^2$. 
For any current sheet to be plasmoid unstable, in MHD regime, the associated Lundquist number has to be larger than a certain critical value \citep{Bhattacharjee_2009}. Thus we require the following constraint to be satisfied, $S_{CS}=B_{rec} l_{CS}/\eta \gtrsim S_{crit}$, where $S_{crit}\sim 10^4$.
Substituting for $\lambda/l_{CS}$ and $S_{CS}$ into \Eq{tratioupd}, we obtain for the MHD case, 
\EQ
\frac{\tau_{acc}}{\tau_{life}} \gtrsim \frac{100}{f_{sat}^3} \left(\frac{B_g}{B_{rec}}\right)^{(s+1)} \left(\frac{S_{crit}}{S}\right)^{s}.
\label{ratio1}
\EN

On the other hand, in a collisionless plasma, the Alfv\'enic timescale is governed by a dispersion 
relation that is different from that of the MHD case and thus we have, $\tau_{rec}=l_{CS}\lambda/v_{A_{CS}}\rho_i$ where $\rho_i$ is the ion gyro-radius. Thus the ratio $\tau_{acc}/\tau_{life} = 10 (l_{CS} v_{CS})/(v_{A_{CS}}\rho_i)$. The important constraint is an empirical condition that has to be met 
for plasmoids to arise, $l_{CS}/\rho_i > (l_{CS}/\rho_i)_{crit}$, where $(l_{CS}/\rho_i)_{crit} = 50$ \citep{Ji_2011}. With this, we obtain, 
\EQ
\frac{\tau_{acc}}{\tau_{life}} \gtrsim \frac{10}{ f_{sat}} \left(\frac{B_g}{B_{rec}}\right) \left(\frac{l_{CS}}{\rho_i}\right)_{crit}^{s+1}  \left(\frac{\rho_i}{L_{out}}\right)^{s}
\label{ratio2}
\EN

Using the equation \Eq{ratio1} and considering the fiducial values for all the parameters i.e. $s=1/3$ (for standard Kolmogorov turbulence), $f_{sat}\sim 0.5$ \citep{Seta_2020}, $\beta_g=100$ and $S=10^{25}$ \citep{Kunz_2011, Kunz_2022}, we obtain the ratio  $\tau_{rec}/\tau_{life} \sim 10^{-5}$ which is reassuring.
We can invert this argument to put bounds on the Lundquist number ratio i.e. for $\tau_{rec}/\tau_{life} < 1$, we have $S_{CS}/S \lesssim 10^{-6}$.
For the collisionless case, in addition to the above mentioned fiducial values, we have $\rho_i \sim 6\times10^8$~m calculated with $B_{rec}$ (we take $B_g=1~\rm{\mu G}$ consistent with $\beta=100$, number density of $10^3~\rm{m^{-3}}$ and $T=10^8~\rm{K}$) and $L_{out}=100$~kpc. Thus the timescales ratio from \Eq{ratio2} is $\tau_{rec}/\tau_{life} \sim  10^{-2}$. Similar to before with MHD case, here we find $l_{CS}/\rho \lesssim 1000$ to satisfy $\tau_{rec}/\tau_{life} < 1$. 

\subsection{Frequency of re-acceleration}
\label{sec:Frequency of re-acceleration}
We now estimate if the magnetic reconnection sites are frequent enough for the re-acceleration to sustain the population of high energy electrons required to maintain the large-scale diffuse nonthermal emission.
 
To estimate the frequency of sites with fields of the strength, $B_{rec}$, consider the distribution function of fields generated by the fluctuation dynamo given by,
\begin{equation}
    f(B) = (2\pi \sigma^2)^{-1/2}\exp\left(\frac{-[\ln(B/B_g)-\mu]^2}{2\sigma^2}\right),
    \label{eq:log_normal}
\end{equation}
where $\mu=-0.29$ and $\sigma=0.71$ are the respective mean and standard deviation of the distribution of $\ln(B/B_g)$ for a system with subsonic turbulence and magnetic Prandtl number $P_M>1$ \citep{Seta_2021}. Note that the PDF is a log-normal distribution, as expected for 
fields of intermittent nature \citep{Zeldovich_1987}. 

From the PDF, we estimate the volume of regions, $\tilde{v}$ that are above the required magnetic field $B_{rec}$. 
The volume of the galaxy cluster under consideration is $V$. The area under the curve of the PDF occupied by fields $\vert B \vert > B_{rec}$ as $r$ and the total area under the curve is $A = 1$. We have, 
$\tilde{v}/{V} = {r}/{A}$ implying $r = \frac{1}{2}\left(1-{\rm erf}\left(({\ln{(B_{rec}/B_g)}-\mu})/{\sqrt{2}\sigma}\right)\right)$.
Hence, $r$ is the volume filling factor. Now, the total volume of all the relevant magnetic structures is given by $\tilde{v}= rV$ and that of an individual magnetic structure hosting a magnetic reconnection site is $v_M=l_{cs}\lambda^2$.  Therefore, the total number of reconnection sites is $N = \tilde{v}/v_M$. Then the distance $l_{rec}$ between two such consecutive sites is given by \EQ
l_{rec} \sim \left(\frac{V}{N}\right)^{1/3} = \left(\frac{v_M}{r}\right)^{1/3}.
\EN
Note that the total volume $V$ does not appear in $l_{rec}$. Nevertheless, the time taken for an accelerated electron with diffusion velocity $v_e \sim 100 \rm{km/s}$, to travel from one site of reconnection to the other is $\tau_{tr}=  {l_{rec}}/{v_e}$. For MHD, we obtain, 
\EQA
\tau_{tr} = \frac{S_{crit}}{S}\frac{L_{out}}{v_e}\frac{1}{\sqrt{\beta}}\left(\frac{f_{sat}^4}{r}\right)^s
\ENA
With the fiducial values, we have, 
\EQA
\tau_{tr} &&\lesssim  2.35\times10^{10}{\rm s}~\left(\frac{S_{crit}/S}{10^{-6}} \right)\left(\frac{L_{out}}{100~{\rm kpc}}\right)\\\nonumber &&\times\left(\frac{100~{\rm km/s}}{v_e}\right)\left(\frac{10}{\sqrt{\beta}}\right)\left(\frac{f_{sat}}{0.5}\right)^{4/3}\left( \frac{1.3\times 10^{-4}}{r} \right)^{1/3}.
\label{eq:interaction_time_MHD}
\ENA
In the collisionless kinetic case, the time taken by the electron to encounter a  reconnection sites is given by,  
\EQA
\tau_{tr} = \left(\frac{l_{cs}}{\rho_i} \right)\frac{\rho_i}{v_e}\left(\frac{f_{sat}}{r}\right)^s. 
\ENA
Again with the fiducial values, we obtain, 
\EQA
\tau_{tr} \lesssim && 6\times10^8 {\rm s}~\left(\frac{l_{cs}/\rho_i} {1000}\right)\left(\frac{\rho_i}{6\times 10^8\ \rm{m}} \right) \\ \nonumber && \times \left(\frac{100\ \rm{km/s}}{v_e} \right)\left(\frac{f_{sat}}{0.5}\right)^{4/3}\left( \frac{1.3\times 10^{-4}}{r} \right)^{1/3}.
\label{eq:interaction_time_kin}
\ENA
 In both the cases, MHD and kinetic, the time taken by the electron to travel between two consecutive reconnection sites is about $10^{10} $ to $10^{8}$ seconds i.e. 1000 to 10 years, which is very small compared to the cooling timescale ($\sim 10^8$ years). 

\subsection{Maximum acceleration}
\label{sec:Maximum acceleration}
Next we would like to calculate the maximum possible particle energisation that can be achieved in a given reconnection site. 
The maximum Lorentz factor is determined by the amount of work done by the electric field in the current sheet, $mc^2(\gamma_{max}-1) = \int eE_{rec}cdt$. Here the $E_{rec}$ is electric field set up by the reconnection, i.e. the rate of change of magnetic flux given by $E_{rec}=\tau_{rec}^{-1}B_{rec}l_{CS}$. Thus, we have,
\EQ
\gamma_{max}\approx \frac{e}{mc}
~\frac{\tau_{life}}{\tau_{rec}}~(B_{rec}l_{CS}).
\label{gammax}
\EN 
To calculate this in the MHD case, we substitute \eq{tratioupd} into \Eq{gammax} to obtain, 
\EQ
\gamma_{max} \sim \frac{e}{mc} \frac{f_{sat}^3}{100} \left(\frac{B_{g}}{B_{rec}}\right)^{-(s+1)} \left(\frac{S_{CS}}{S}\right)^{1-s} B_g L_{out}.
\label{maxgamma1}
\EN
Again considering the fiducial values, we have, 
\EQA
\gamma_{max} && \sim 2.5\times(10^{-2}-10^8) \left(\frac{S_{CS}/S}{10^{-21}-10^{-6}}\right)^{2/3} \nonumber \\ 
&& \times \left(\frac{B_g}{1 \rm{\mu G}}\right) \left(\frac{L_{out}}{100 \rm{kpc}}\right).
\label{maxgammaval1}
\ENA
Here, $\gamma_{max}$ has a range from $10^{-2}$ to $10^{8}$ depending on the Lundquist number values. It depends crucially on the ratio of $S_{CS}/S$. The value of $S\sim10^{25}$ is from a Spitzer estimate and hence is likely exaggerated. With lower values of $S$ or higher values of $S_{CS}/S$, we achieve higher $\gamma_{max}$. The upper limit of $(S_{CS}/S) \lesssim 10^{-6}$ leads to $\gamma_{max} < 10^8$.
In the collisionless kinetic case, we have,
\EQA
\gamma_{max} \sim && \frac{e}{mc} \frac{f_{sat}}{10} \left(\frac{l_{CS}}{\rho_i}\right)^{-s}  \left(\frac{\rho_i}{L_{out}}\right)^{-s+1} 
\nonumber \\ 
&& \times \left(\frac{B_{rec}}{B_g}\right)^2 B_g L_{out}
\label{maxgamma2}
\ENA
With the fiducial values, for the estimate for maximum Lorentz factor, we have, 
\EQA
\gamma_{max} \sim && (1.7-4.25)\times 10^5  \left(\frac{(1/1000) - (1/50)}{\rho_i/l_{CS}}\right)^{-1/3} \nonumber \\
&& \times \left(\frac{\beta_g}{100}\right)\left(\frac{B_g}{1~\rm{\mu G}}\right) \left(\frac{L_{out}}{100~\rm{kpc}}\right).
\label{maxgammaval2}
\ENA
Thus, we find that the range of $\gamma_{max}$ is much tighter than that in the MHD case. The value of $10^5$ is also the upper limit of the range in $\gamma$ inferred from observations from radio relics. Thus, these estimates, which align with observations, make acceleration due to reconnection a promising model. 

\subsection{Radio Luminosity} 
With our model, we calculate here the radio luminosity from the relativistic electrons. The power radiated by a single electron of Lorentz factor $\gamma$ in the magnetic field $B$ is given by 
\begin{equation}
    P_{syn}(\gamma)\simeq\frac{4}{3}\sigma_T c U_{mag}\gamma^2 = \frac{4}{3}\sigma_T c \frac{B^2\gamma^2}{2\mu_0},
    \label{eq:syn_power_single_e}
\end{equation}
where, $\sigma_T$ is Thomson scattering cross section, $c$ is the light speed and $U_{mag} = B^2/2\mu_0$. However, to obtain the power radiated by the electron population with a number density $N(\gamma) = N_0\gamma^{-p}$, {we invoke the equipartition between particles’ and magnetic field’s energy, i.e., 
\EQA
{\int_{\gamma_{min, N_0}}^{\gamma_{max, N_0}} d\gamma} N(\gamma) \gamma mc^2 = f_{eq}\frac{B^2}{2\mu_0}.
\label{equpart}
\ENA
where $f_{eq}\sim 0.25~-~0.5$ is the fraction of energy in the ultra-relativistic electrons.}
Thus, the bolometric synchrotron luminosity is given by, 
\EQA
L_{syn} = && V \int_{\gamma_{min}}^{\gamma_{max}} d\gamma P_{syn}(\gamma)N(\gamma)
\\
\nonumber = && \left(\frac{4}{3}\frac{\sigma_Tc}{(2\mu_0)^2 mc^2}\right) \left(f_{eq}VB^4 \right) \left(\frac{2-p}{3-p}\right)\\
\nonumber && \times \left(\frac{\gamma_{max}^{(3-p)}-\gamma_{min}^{(3-p)}}{\gamma_{max, N_0}^{(2-p)}-\gamma_{min, N_0}^{(2-p)}}\right), 
\label{eq:L_syn_N_gamma}
\ENA
where $V$ is the volume of cluster with 1Mpc radius and the normalization constant is from \Eq{equpart},  \EQ
N_0 = f_{eq}\frac{B^2(2-p)}{2\mu_0 mc^2 \left(\gamma_{max, N_0}^{(2-p)}-\gamma_{min, N_0}^{(2-p)}.\right)}
\EN

Considering all the relativistic electrons, the limit of the integration are $\gamma_{min, N_0} = 2$ and $\gamma_{max, N_0} = 10^5$.

Now, we use $p = 3.8$, $\gmx = 10^5,\ \gmn = 10^3$ (relevant for the radio), $B = B_{rec} = 10B_g = 10\ \mu\rm{Gauss} = 10^{-9}\ \rm{Tesla}$, $f_{eq}=0.5$ and $V\simeq 4\times(1\rm{Mpc})^3 = 4\times3^3\times 10^{66}\ \rm{m}^3$ in equation (\ref{eq:L_syn_N_gamma}) and obtain,
\begin{eqnarray}
L_{syn} && = 0.6 \times10^{35}\times\lb\frac{\sigma_T}{6\times10^{-29}}\rb \lb\frac{c}{3\times10^8}\rb \\ \nonumber \times && \lb\frac{4\times1.44\times 10^{-12}}{4\mu_0^2}\rb\lb\frac{10^{-13}}{mc^2}\rb \lb\frac{V}{4\times3^3\times 10^{66}}\rb \\ \nonumber && \times \lb\frac{B^4}{10^{-36}}\rb\lb\frac{-0.8}{3-p}\rb\lb\frac{2-p}{-1.8} \times\lb\frac{\gmx^{(3-p)}-\gmn^{(3-p)}}{10^{-4}-10^{-2.4}}\rb \rb\\
\nonumber && \lb\frac{10^{-9}-2^{-1.8}}{\gmxp^{(2-p)}-\gmnp^{(2-p)}}\rb\ \rm{Joules/s}. \end{eqnarray}
 Hence, the calculated radio luminosity is $6\times 10^{41}$ ergs/s, which is in agreement with the observation \citep{Feretti_2012, vanWeeren_2019}. 
 
{ One could include a volume filling factor of the reconnecting current sheets in the calculation above, which we estimated to be around $10^{-4}$ in section~\ref{sec:Frequency of re-acceleration}. However, we believe the effective volume filling factor would be much larger due to two reasons. Firstly, the field is extremely dynamical which can result in reconnection sites appearing at different locations from one moment in time to another, this time scale has to be necessarily  smaller than the eddy turnover time scale considering that the smallest length scale is the current sheet width or the very sharp gradients in the magnetic field.
Secondly, the electrons which get accelerated also spread out and encounter another reconnection site much before they cool down as given by the ratio of the travel time to cooling time in our paper. In this way, the effective volume filling factor could be much larger, closer to $0.01$ or $0.1$.

{We saw earlier from \Fig{fig:energy_spectra_theta_point05_diff_boxsizes_inset}, that $p=3.8$ is not necessarily the final slope at $\theta=0.05$ and it does get smaller with increasing box sizes and ultimately the smaller slope value could compensate the decrease due to the filling factor.}}

\section{Conclusions}
\label{sec:Discussion and Conclusion}
In this work, we provide an alternative explanation for  the origin of nonthermal emission in the halo of galaxy clusters. Due to the predominantly turbulent nature of the ICM, the operating fluctuating dynamo naturally generates sites of magnetic reconnection with reversing magnetic fields.
Using PIC simulations (given that ICM is largely collisionless) for nonrelatistic plasmas, we find that the reconnection process accelerates the thermal electrons in the non-relativistic regime to become relativistic and non-thermal. 
{The slope of the non-thermal electron energy spectrum for the fiducial case is within the observational range of flatter spectral indices. As the temperature is increased by even a small amount, the slope of the spectra become much shallower.} {The schematic diagram of how energy gets transferred from turbulence to electrons via reconnections and hence corresponding radio emission via magnetic reconnection according to our model is shown in Figure~\ref{fig:bigger_picture}. }

To check the feasibility of the particle acceleration process via reconnection in the context of galaxy clusters, we also estimate different timescales: namely, the timescale on which accelerated electrons encounter  reconnection sites, acceleration timescale and eddy turnover time in both MHD and kinetic regimes. 
Our estimates indicate that the accelerated electrons encounter reconnection sites few orders of magnitude faster than their cooling timescale. Thus, it supports the extended radio emission observed in the radio halo. 

For efficient particle acceleration, the reconnetion rate should be faster than eddy turn over rate. We, therefore, estimate the ratio between these two timescales in both, MHD and kinetic regime.   
In both cases, we find the ratio to be substantially  small. 

Another important quantity we estimate is the maximum energy ($\gamma_{max}$) of the accelerated electrons.  We find that the MHD calculation results in a less stringent range of $\gamma_{max}$, 
nevertheless yielding very large values upto $10^8$ for larger current sheets. 
The collisionless calculation, on the other hand, yields a tighter range around the value of $\sim 10^5$ that well aligns with observations.
{The radio luminosity estimate from our model is also in agreement with the observation.}
In conclusion, this letter demonstrates that magnetic reconnection arising from the fluctuation dynamo is a viable mechanism for particle energization  since it can explain three key observational properties of the radio halo: the observed radio spectral indices, the maximum energy of the radio halo electrons, {and the radio luminosity in the radio halo}.
{While this letter demonstrates the possibility of magnetic reconnection as a viable mechanism for acceleration in galaxy clusters, detailed comparison with observation and further rigorous calculations pertaining to acceleration modeling are left to the future.}


\section{Acknowledgment}
We thank Kandaswamy Subramanian for very useful comments on the paper and Radhika Achikanath Chirakkara for discussions on PIC simulations early in the project. 
{We would like to thank the reviewer for their comments and suggestions, that helped improving the readability of this paper.} We also acknowledge discussions and talk recordings from the forum of Women in Astrophysical Fluids and Plasmas (http://womeninastro-fluids-plasmas.org). Similarly with NORDITA dynamo seminars. 
S. G. would like to thank Chandranathan Anandavijayan for going through the manuscript and giving important feedback. {S. G. acknowledges the Knowledge Exchange grant, which sponsored his travel to the conference on Galaxy Clusters \& Radio Relics II, in Cambridge, MA. S. G. thanks the organizers for supporting his stay. S. G. had illuminating discussions that helped to improve the paper.} All the simulations were performed in the Contra cluster at the International Centre for theoretical sciences. 
We acknowledge support of the Department of Atomic Energy, Government of India, under project no. RTI4001.

\appendix

\section{Particle acceleration mechanism}
\label{sec:Particle_acceleration_mechanism}

\begin{figure}
    \centering
    \includegraphics[width=\columnwidth]{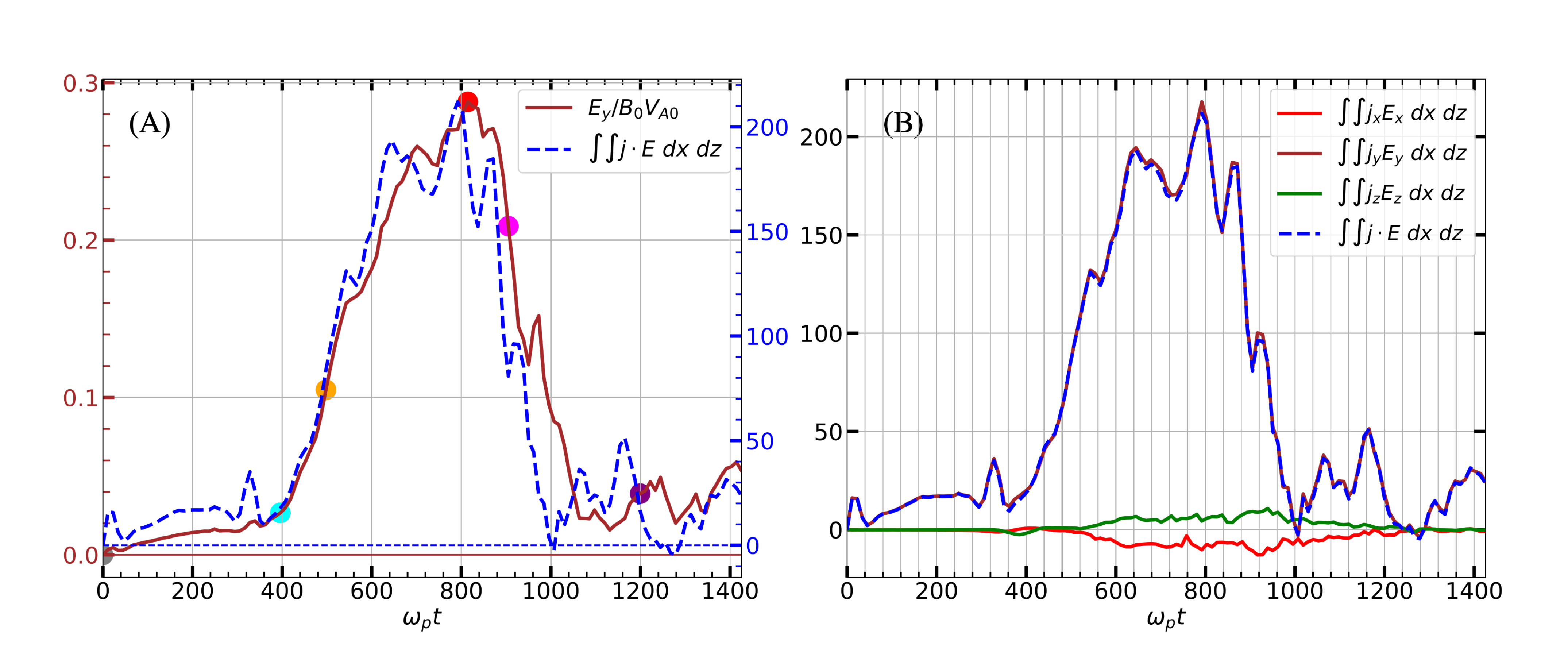}
    \caption{(A) shows the evolution of the reconnection electric field, $E_{rec}=E_y$ (the brown solid line), and the rate of energy gain by the electrons, $\int \int \tbf{j}\cdot \tbf{E}\ dxdz$ (the blue dashed line) for the case of $\theta = 0.05$. The horizontal lines indicate the corresponding zero values on the ordinate axes. $E_{rec}$ is normalized by $B_0V_{A0}$, where $V_{A0} = B/\sqrt{2\mu_0 m n_0}$ is the Alfv$\acute{e}$n velocity. (B) shows the evolution of the contributions to $\int \int \tbf{j}\cdot \tbf{E}\ dx dz$ from different components of the electric field, from our fiducial run with $\theta=0.05$.}
    \label{fig:E_field_jdotE_comp}
\end{figure}
{
In our simulations, the background electrons are  accelerated mainly via reconnection electric field, as would be the case when there is no guide field \citep{BB_2007}. 
To gain insight into the driving mechanism of particle acceleration, we study the evolution of the rate of energy gain by the electrons, i.e., $W_e = \jdotE$. 
The reconnection electric field, $\tbf{E}_{rec}=E_y \boldsymbol{\hat{\tbf{y}}}$ has only $y$-component as expected from our setup. \Fig{fig:E_field_jdotE_comp}(A) shows the evolution of $W_e$. 
We find that its temporal evolution has a correlation with that of the reconnection electric field, $E_{rec}$, which is over-plotted in the same figure. 
For further confirmation, in \Fig{fig:E_field_jdotE_comp}(B), we study the contributions to $W_e$ from different components of the electric field, i.e., $W_e = \sum_{i = x, y, z}W_{e,i}$, where $W_{e,i} = \int \int j_i E_i\ dx dz$. The $\int \int j_y E_y\ dx dz$, due to the reconnection electric field, is the dominant one among the three and it alone traces the evolution of $W_e$. That is why we claim that the reconnection electric field, $E_y$ is the main accelerating agent in our setup. Further details of acceleration will be delineated in the next paper.}

\bibliography{sample}{}
\bibliographystyle{aasjournal}



\end{document}